\documentclass{aastex}
\usepackage{emulateapj5}

\shorttitle{CTB104A: Magnetic Field \& Environment}
\shortauthors{UYANIKER et al.}

\begin{document}

\title{The Supernova Remnant CTB104A : Magnetic Field Structure and
Interaction with the Environment }
\author{B\"ULENT  UYANIKER\altaffilmark{1},
        ROLAND  KOTHES\altaffilmark{1},
 and CHRISTOPHER M. BRUNT\altaffilmark{1,2}}
\altaffiltext{1}{National Research Council, 
       Herzberg Institute of Astrophysics, 
       Dominion Radio Astrophysical Observatory, 
       P.O. Box 248, Penticton, B.C., 
       V2A 6K3 Canada
    }
\altaffiltext{2}{
         Department of Physics and Astronomy, 
         The University of Calgary, 
         2500 University Dr. NW, 
         Calgary, AB, 
         T2N 1N4 Canada
 }

\email{bulent.uyaniker@nrc.ca,
roland.kothes@nrc.ca,
chris.brunt@nrc.ca
}

\begin{abstract}
We present new, high resolution 1420 and 408~MHz continuum images and
\ion{H}{1} and $^{12}$CO (J=1$-$0) spectral line maps of the diffuse
supernova remnant CTB104A (G93.7$-$0.3). Analysis of the complex
continuum emission reveals no significant spectral index variations
across the remnant. Three prominences around CTB104A are found to be
related to the SNR, while one extension to the east is identified as
an \ion{H}{2} region associated with a background molecular shell.
Small scale polarization and rotation measure (RM) structures are
turbulent in nature, but we find a well-ordered RM gradient across the
remnant, extending from southeast to northwest.  This gradient does
not agree with the direction of the global Galactic magnetic field,
but does agree with a large-scale RM anomaly inferred from rotation
measure data by \citet{clegg}.  We show that the observed morphology
of CTB104A is consistent with expansion in a uniform magnetic field,
and this is supported by the observed RM distribution.  By modeling
the RM gradient with a simple compression model we have determined the
magnetic field strength within the remnant as $B_{0}\approx 2.3~\mu$G.
We have identified signatures of the interaction of CTB104A with the
surrounding neutral material, and determined its distance, from the
kinematics of the \ion{H}{1} structure encompassing the radio
emission, as 1.5 kpc.  We also observed clear breaks in the \ion{H}{1}
shell that correspond well to the positions of two of the prominences,
indicating regions where hot gas is escaping from the interior of the SNR.
\end{abstract}


\section{Introduction}
CTB104A (G93.7$-$0.3) is a diffuse, thick-shelled supernova remnant
(SNR) with four ``prominences'', extending beyond the main body of the
shell.  Three of the prominences are polarized, as reported by
Mantovani et al. (1991).  The radio spectrum of CTB104A (S$_{\nu}$
$\propto$ $\nu^{-0.42}$) is relatively flat, typical of a mature
shell-type SNR in the radiative expansion phase.

In broad terms, CTB104A appears as a double limb-brightened shell.
The lack of symmetry between the limbs may be indicative of different
shock conditions across the remnant and hence a range of effective
evolutionary ages.  \citet{kesteven} interpret CTB104A as a possible
barrel-shaped SNR with the shell of the remnant being physically
distorted by the interstellar medium and elongated in the direction of
the Galactic magnetic field.  This is consistent with the observed
morphology of CTB104A, but only for a particular choice of magnetic
axis.  Consequently, good quality polarization measurements can be
used to examine this possibility.

Apart from identifying the origin of the shell morphology, a complete
understanding of CTB104A depends upon gaining information on the
nature of the prominences extending from the shell.  With the
exception of the prominence to the east (higher longitude) the
prominences do not terminate at a sharp boundary, but rather appear to
fade smoothly into the Galactic background emission.  These
prominences may be ``hot blasts'' escaping into regions where the
density of the environment is presumably relatively low, as suggested
by \citet{landecker85}.  A close examination of the surrounding
neutral medium may reveal signatures relating to the origin of the
prominences.

To diagnose the dominant processes that shape CTB104A and to gain
insight on the prominences extending from the remnant, we have
compiled high resolution radio continuum data at 1420 and 408~MHz,
polarization measurements at 1420~MHz, \ion{H}{1} spectral line data
and HIRES processed IRAS images \citep{cao} from the Canadian Galactic
Plane Survey (CGPS), as well as $^{12}$CO (J=1$-$0) emission line data
at comparable resolution from the Five College Radio Astronomy
Observatory (FCRAO).  We examine the spectral index distribution with
TT-plot analysis between 408 and 1420~MHz for signs of differing shock
conditions over the remnant's shell and also towards the prominences.
Linear polarization measurements are used to study the overall role of
the magnetic field on the expansion of the SNR and to search for
spatially variable polarization that is indicative of regions of
turbulence and material heated by localized shock compression.  We
identify signatures of the interaction of the SNR with its
surroundings from the CGPS \ion{H}{1} line data and use this
additionally to constrain the distance to CTB104A as $\sim1.5$~kpc.
Finally, we use the FCRAO CO data and HIRES IRAS data to identify one
of the prominences as a background \ion{H}{2} region, located in the
Perseus Arm, and thus not associated with the SNR.

\section{Observations}
\subsection{CGPS data}
CTB104A was observed at 1420 and 408~MHz in continuum mode, including
the Stokes Q, U and V parameters at 1420~MHz.  Radio continuum and
\ion{H}{1} line observations were carried out simultaneously  with the
Dominion Radio Astrophysical Observatory (DRAO) synthesis telescope
\citep{landecker00} as part of the CGPS. HIRES-processed IRAS data
\citep{cao} were also obtained from the CGPS data archive.  The CGPS is
described by \cite{taylorcgps} and a more detailed description of the
data processing routines can be found in \cite{willis}.  The angular
resolution of the DRAO synthesis data varies slightly across the final
maps as
$1\arcmin\times1\arcmin$~cosec($\delta$)(1420/$\nu$(GHz)).
Single antenna data are incorporated into the synthesis maps to assure
accurate representation of all structures up to the largest
scales. The low spatial frequency continuum data are obtained from the
Effelsberg Survey at 1420~MHz \citep{reich} and at 408~MHz from Haslam
et al. (1982). The low spatial frequency \ion{H}{1} data are from the
Low Resolution DRAO Survey of the CGPS region observed with the DRAO
26-m Telescope \citep{higgs}.

\subsection{FCRAO CO data}

We have obtained spectral line images of $^{12}$CO (J=1$-$0) emission in
the area around CTB104A from the FCRAO 14m telescope. Our mapped area
covers Galactic longitudes $\ell\sim93\degr$ to $\ell\sim94\fdg9$ and
Galactic latitudes $b\sim-1\degr$ to $b\sim+0\fdg33$, with a somewhat
irregular boundary.  The CO data for $\ell<94\fdg5$ were obtained by
Routledge and Moriarty-Schieven (unpublished) in November-December
1994 using the QUARRY 15 pixel focal plane array \citep{erickson1992};
to this we have added more data at $\ell>94\fdg5$, taken during
March/April 2001, using the SEQUOIA 16 pixel focal plane array
\citep{erickson1999}. Pointing measurements were carried out using
Venus (1994 data) and the SiO maser $\chi$ Cygni (2001 data).  Both
sets of CO data were obtained by position-switching.

 The 1994 data have 80~MHz ($\sim$190 km~s$^{-1}$) total bandwidth,
with the spectrometer centered on $v_{\rm LSR} = -45$ km~s$^{-1}$ and
the 2001 data have 40~MHz ($\sim$95 km~s$^{-1}$) total bandwidth
centered on $v_{\rm LSR}=-30$ km~s$^{-1}$.  The spatial sampling and
velocity resolution (1.21 times the channel spacing) for the 1994 and
2001 data sets respectively were 25$\arcsec$, 0.98 km~s$^{-1}$ and
22$\arcsec$, 0.25 km~s$^{-1}$.  The spatial resolution (beam FWHM) of
the data is 45$\arcsec$.  Both data sets were initially calibrated by
the chopper wheel method \citep{kutner} and subsequently converted to
the radiation temperature (T$_{R}^{*}$) scale by correcting for
forward scattering and spillover losses ($\eta_{fss}$ = 0.7).  Each
data set was convolved to 2$\arcmin$ spatial resolution and 0.98
km~s$^{-1}$ velocity resolution prior to combining them on the same
$\ell, b, v_{\rm LSR}$ grid as the \ion{H}{1} data.  The resulting
sensitivity was 0.29~K and 0.08~K (1$\sigma$) for the 1994 and 2001
data sets respectively.
 
\subsection{Morphology of the SNR} 

The radio continuum total intensity maps around CTB104A are shown in
Figures~\ref{i1420} and \ref{i408}, at 1420 and 408~MHz respectively.
The remnant has a roughly circular shape consisting of three distinct
fragments. A circle of $1\fdg 2 \times 1\fdg 2$ centered at
$\ell=93\fdg7$ and $b=-0\fdg3$, fitted by eye and shown in
Figure~\ref{finder}, delineates this roughly circular shape.  The
prominences, identified in Figure~\ref{finder} by arrows, form an irregular
boundary to the otherwise circular remnant.  The angular size of the
SNR, taking all the prominences into account, is about $1\fdg
3\times1\fdg 5$.  The northern part of the shell has the highest
intensity at both 408 and 1420~MHz maps.  The prominence extending
toward the south of the SNR, where the shell is incomplete, is a good
candidate for an example of the ``break out'' phenomenon observed in
mature SNRs (e.g. Cygnus Loop, VRO~42.05.01 and  CTA~1). A
similar claim could be made for the other prominences if projection
effects are considered.

Disregarding the prominences, we examine the idea that the SNR's
morphology could be interpreted as due to the expansion of the remnant
in a regular magnetic field.  In such a model (e.g. van der Laan 1962)
no emission in total-intensity along the field direction is expected
if the surrounding magnetic field is uniform on a scale larger than
the SNR.  The positions of the three shell fragments can support this
model only for a particular choice of magnetic field axis. This
preferred axis, lying $\sim$ 20 - 40 degrees north of west
\footnote{20 - 40 degrees counter clock wise with respect to the
Galactic plane.},  partitions
the remnant into two halves, one consisting of the north and east
fragments and the other consisting of the southwest fragment alone. No
other choice of magnetic field axis can validate this model.  In
Section~2.5 we show that the actual magnetic field axis lies at $\sim$
30$\degr$ north of west.  Thus, expansion in a uniform magnetic field
is consistent with the observed morphology of the SNR.

The diffuse filamentary structure with a bright core seen to the south
of the SNR is the \ion{H}{2} region S124.  The distance of S124 based
on the distance of the exciting star from spectrometric data is
$d=2.6\pm0.6$ kpc (Felli \& Harten 1981, Brand \& Blitz 1993), which
excludes a physical relationship between S124 and the remnant since
CTB104A is at a distance of $\sim$1.5 kpc (see Section~\ref{hi}).

\subsection{Spectral analysis} \label{specan}
Spectral indices were computed using the conventional TT-plot method
by plotting the brightness temperature values of the 408~MHz maps
against those at 1420~MHz. The 1420~MHz data are convolved to the
resolution of the 408~MHz map.  To avoid oversampling
in the TT analysis the maps are regridded to 1$\arcmin$. 
In order to reveal any variation in the spectral index across the
whole SNR, we have selected regions towards distinctly separable parts
of the remnant and performed TT-plot analysis on the data extracted
from those individual regions.  Regions corresponding to the core of
S124 and the extragalactic source 4C50.55 were also included in this
analysis for a reliability check on our measured indices.
Figure~\ref{finder} shows the selected regions marked on the 1420~MHz
image.  The regions were selected in a way to avoid unresolved
sources, but this was not always possible. Since these sources are
most likely extragalactic, we have checked the images for unresolved
sources in the highest resolution image (1420~MHz) and we excluded a
circular region with a diameter of 4$\arcmin$ around these sources.
Then we plotted brightness temperature values of both maps with
respect to each other and fitted the distribution with a straight
line. The slope of this line gives the temperature spectral index
$\beta$, where T$_{\rm B}\sim\nu^{\beta} = \nu^{\alpha -2}$.  We
repeated this analysis by changing the independent variable and obtain
another value for $\beta$. We then adopted the difference of these two
values of $\beta$ as the error in the spectral index. Spectral indices
derived this way are not strongly affected by the presence of unknown
backgrounds contributing to the emission brightness at each
frequency. The presence of uniform backgrounds will modify only the
unimportant constant term in the fit, leaving the derived slope
unaffected, while varying backgrounds will tend to increase the
scatter.

The results of the TT analysis are plotted in Figure~\ref{spec}.  The
analysis gives $\alpha=-0.1$ for S124, which is typical for an
optically thick \ion{H}{2} region. We obtain $\alpha=-0.59$ for
4C50.55, in agreement with the value given by Mantovani et al. (1982)
obtained from integrated flux densities compiled from the literature
between 178~MHz to 10.7~GHz. Spectral indices generally show a
relatively flat spectrum for the SNR and within the errors of the
measurements we do not find any variations in the spectral index
across the remnant, except for the central region (s4), where we
observe a steeper spectrum. However, in the central region the surface
brightness is very low, giving rise to high uncertainty.  

The boxes e1, e2, e3 ,e4 are positioned to sample the spectral index
of the prominences (see Fig. \ref{finder} and Fig. \ref{spec2}).  The
general trend is that spectral indices towards the prominences are
similar to those seen in the main body of the SNR's shell, albeit with
larger errors arising from the weakness of the prominence emission,
except for the region e2 which indicates a thermal spectrum. We
interpret the spectral index measurements as supporting evidence that
three of the prominences indeed are non-thermal and related to the
SNR, as suggested by the overall SNR morphology.  The region e2,
however, seems to be an unrelated source of thermal origin, and indeed
this prominence is unpolarized (see Mantovani et al. (1991) and
Section~\ref{pol} below). For further evidence of the association of
the remaining three prominences to the SNR we must appeal to the
polarization measurements discussed below.

The lower-right panel of Figure~\ref{spec} shows a TT-plot using the
region indicated by dotted lines in Figure~\ref{finder}. From this
region we derive a global spectral index of the SNR as $\alpha
\sim-0.43$. This calculated spectrum is comparable to the value
$\alpha=-$0.42 given by \citet{mantovani91}, but still  flatter
than the typical value of $\alpha=-$0.5 for an SNR expanding
adiabatically.  This indicates that the SNR is most likely already in
the radiative phase.

\subsection{Polarization characteristics of the SNR}\label{pol}

Noise corrected polarized intensity (PI) maps are calculated as
$PI=\sqrt{ U^2+Q^2-(1.2~\sigma)^{2} }$, where $\sigma=30$ mK T$_{\rm B}$
is the rms noise in each map.  Polarization angle maps have been
derived, as $\psi_{\lambda} = \case{1}{2} ~\arctan \case{U}{Q}$, from
the four bands of the DRAO telescopes, each of width 7.5~MHz and
centered at 1406.65~MHz (band A), 1414.15~MHz (B), 1426.65~MHz (C) and
1434.15~MHz (D). These maps have been used to calculate the rotation
measure across the remnant.  In Figure~\ref{pirm} we show the
polarized intensity and rotation measure maps of CTB104A.  We observe
polarization from almost all parts of the remnant.  A distinct
difference between the total intensity and the polarized intensity
images is immediately evident: there are more small-scale structures
in the polarized intensity image than in the total intensity image.
This is the result of high rotation measure differences on small
scales within the remnant, which we assign to turbulence and/or
localized shock compression.

The northern part of the remnant is particularly affected by
depolarization; a distinct region of low polarized intensity can be
seen in this area. Below this area, we observe patchy polarization
structures.  In the polarization image it is also possible to see
filamentary regions where PI is zero.  Rather than signalling the
absence of polarization, these structures are most likely due to beam
depolarization, arising from a fluctuating magneto-ionic medium within
the beam.

The percent polarization varies from 5 to 20\% over the remnant.  The
extension at the western edge of the SNR is weakly polarized.  There
is polarization towards all prominences, except the one in the
east. Given the measured spectral indices of the prominences
(consistent with a non-thermal spectrum and also similar to the
indices in the SNR shell) along with their observed polarization, we
conclude that with the exception of the eastern one, the prominences
are associated with CTB104A.

Towards CTB104A there is smooth detectable total intensity emission,
and yet the detected polarization is dominated by small-scale
structures.  Bandwidth depolarization can not be the cause, because
the four individual bands allow us to calculate RM values as high as
$\sim$6000 rad m$^{-2}$. The spatial filtering of the smooth
structures beyond the limited scale of the interferometer affects both
the total intensity and the polarization channels in the same
way. However, polarized intensity is additionally affected by the
foreground Faraday rotation on various scales, even if the intrinsic
polarization is smooth.

\citet{clegg} observed RMs from 54 extragalactic sources and reported
an anomalous RM region, $\ga1000$ rad m$^{-2}$, towards ($\ell,
b)\simeq(92\degr, 0\degr$).  The angular extent of this region is
coarsely constrained by point source probes to be between $1\degr$ and
$8\degr$, indicating enhanced RMs behind the Cygnus region.  By
looking at the RM differences of the double-lobe sources they conclude
that RM differences are dominated by small-scale structures. The
prevalence of small-scale variability is obvious in Figure~\ref{pirm},
but the RM structure also has a systematic component --- a gradual
variation of the RMs in the direction of south-east to north-west,
which is very likely related to the structure seen by \citet{clegg}.
The direction of maximal gradient in RM lies between 30 $\pm$ 10
degrees north of west, although this is not easily seen in a gray scale
image.

In order to exhibit this gradient more effectively we project the RM
distribution on to an axis 30$\degr$ north of west and show the
resulting variation of RM along this axis in Figure~\ref{rmplot}.  The
RM measurements show a systematic increase from negative to positive
with respect to the position along this axis.  We identify this
30$\degr$ axis as the magnetic field axis in the remnant. The observed
orientation of the magnetic field axis supports the idea that the
shell configuration in CTB104A arises from expansion in a uniform
magnetic field.  Indeed, this axis, within the precision to which it
can be ascertained, is the only one that can validate this model for
CTB104A.  The RM gradient indicates that the magnetic field is
tangential across the remnant, which is an accepted field
configuration for relatively old SNRs.

This RM variation can not be attributed to a foreground Faraday
screen, since the required mean electron density would be
$\sim$0.2 cm$^{-3}$ (for an assumed magnetic field strength of
1~$\mu$G  over a pathlength
of 1.5 kpc) which is extremely  high by typical Galactic
standards and should be observable in emission. 
Further, a local structure capable of producing the RM
variation would have to have the exact shape and location as CTB104A,
which is unlikely.  This magnetic structure is thus intrinsic to
CTB104A or to its vicinity.

\section{Neutral Material near CTB104A} \label{hi}

To search for signatures of interaction of CTB104A with the
surrounding neutral material we now examine the CGPS \ion{H}{1} data.
The final \ion{H}{1} mosaic consists of 9 CGPS fields, which we have
convolved to 2$\arcmin$ resolution.  We have integrated the maps at 2
km~s$^{-1}$ intervals and have removed the large scale background
emission from each channel using the BGF algorithm by
\citet{sofue}. These maps are plotted in Figures~\ref{h1} and
\ref{h1b}.
 
Starting from negative velocities we see emission surrounding the
remnant, but the most important concentration of \ion{H}{1} is around
$-$2 to $-$10 km s$^{-1}$, forming a D-shaped structure surrounding
the SNR.  This structure slowly fades out with increasing negative
velocities.  The spatial coincidence of the \ion{H}{1} with the
continuum emission is more prominent between $-$6 and $-$8
km~s$^{-1}$.

Between 10 and 8 km~s$^{-1}$ there is an irregular patch of \ion{H}{1}
emission centered on the SNR, which we tentatively interpret as the
cap of the \ion{H}{1} bubble moving away from us. The approaching cap
is not visible, probably due to confusion with material from the local
spiral arm.

As a modifier to the overall shell morphology of the neutral material
surrounding CTB104A, there are clear breaks in the shell that
correspond well to the positions of two of the prominences.
Particularly striking is a large break occurring near the location of
the northwest prominence at LSR velocities between $\sim-14$ and $-20$
km~s$^{-1}$. There is a plume of neutral material (at $\ell \sim
92\fdg6$, $b\sim 0\fdg6$) which appears to extend beyond, and along a
similar axis to, the prominence seen at 1420~MHz. In this same
velocity range, the neutral material is evacuated from, and appears to
encompass, the region into which the southern prominence
extends. Towards the remaining, smaller, northeastern prominence
associated with the SNR we see no clear signs of interaction with the
surrounding neutral material. However, at the location of this
prominence, the upper radio shell is broken in two; this suggests that
there is a similar escape phenomenon occurring here, but that the
complexity of the \ion{H}{1} emission on small scales precludes direct
identification of interaction of the SNR with its surroundings.  We
thus interpret the prominences as hot gas from the interior of the SNR
escaping into less dense regions, as suggested by \citet{landecker85}.

We identify the surrounding \ion{H}{1} structure at $\sim-6$
km~s$^{-1}$ as defining the rest velocity (relative to the LSR) of
CTB104A, which places the remnant at a distance of $\sim1.5\pm0.2$
kpc, assuming a standard flat Galactic rotation curve (R$_\sun$ =
8.5~kpc, V$_\sun$ = 220 km~s$^{-1}$).  At this distance the linear
size of the remnant is $\sim$35~pc.  Note that, at lower negative
velocities the structure surrounding the northern part of the SNR is
more prominent while the southern part is best visible at higher
negative velocities.  This indicates that the northern part of the
bubble is moving away from us while the southern part is approaching
us. Additionally, the radial velocity of the cap at $\sim$10
km~s$^{-1}$ indicates an expansion velocity of about 16~km~s$^{-1}$
for the bubble.

We also examined our CO-data for a possible correlation with the SNR.
But we could not find any structure which might be related. Further
observations of the surroundings and the interior of the SNR are
necessary, since our measurements only covered the shell of the SNR
and the eastern \ion{H}{2} region.

\section{The detected \ion{H}{2} region G94.48-0.3}

Towards $\ell=94\fdg48$ and $b=0\fdg3$ one of the prominences is
visible in both 408 and 1420~MHz images. In contrast to the other
three structures, we observe no polarization towards this extension of
the remnant. Due to the low intensity values and flat spectral index
of the SNR, it is not possible to determine whether this prominence is
thermal or non-thermal.  However, examination of infrared data shows
that this structure is seen as a prominent half-shell in 60 $\mu$m
IRAS emission (see Figure~\ref{hii}).  This shell is also identified
in our CO line data.  We have integrated the CO emission over the
velocity interval $-55$~km~s$^{-1}$ to $-35$~km~s$^{-1}$, over which
the CO morphology is similar to that of the shell seen at 60~$\mu$m
(see Fig.~\ref{hii}).  The coincidence of molecular emission, infrared
emission and radio emission with a broadly thermal spectrum strongly
suggests that the shell is an \ion{H}{2} region. Moreover, the
observed LSR velocity of the CO emission indicates that the shell has
a distance of $\sim$4~kpc.  Therefore this previously unknown
\ion{H}{2} region is located behind the supernova remnant, which we
expected since the polarization images reveal no additional
depolarization of the supernova remnant in the direction of the
\ion{H}{2} region.

\section{Discussion}

CTB104A is striking due to its shape, but more importantly due to its
distinct rotation measure gradient.  Such a gradient of RM, with a
clear symmetry on both sides of the SNR, signifies a tangential
magnetic field distribution in the remnant. The shock of the explosion
compresses and deforms the magnetic field and thus the remnant carries
the signature of the local magnetic field.  In order to interpret this
RM gradient in terms of the expansion of the SNR in a uniform magnetic
field, we need to have information about the large scale magnetic
field structure towards the SNR. The current method to determine the
direction of the magnetic field in the Galaxy is the use of pulsar
rotation measures (see Han et al. 1999), because they are relatively
easy to measure and in most cases distances of the pulsars are known.
A plot showing the magnetic field direction with respect to the spiral
arms of the Galaxy is given in Figure~\ref{galrm}.  According to this
figure, the direction of the ambient magnetic field is pointing away
from us and directed towards lower longitudes.

Assuming CTB104A is expanding in this uniform large-scale magnetic
field configuration, it should form a bubble by stretching the
magnetic field lines as outlined in Figure~\ref{sketch}.  Such a
magnetic bubble would result in a decreasing RM with respect to
Galactic longitude (left panel in Figure~\ref{sketch}).  However the
observed RM gradient decreases in the opposite direction, namely, the
RM is increasing with longitude.  This implies an ambient magnetic
field direction opposite to the overall Galactic magnetic field.  Thus
the orientation of the magnetic field in this region is not parallel
to the galactic plane, nor in agreement with the ``large-scale''
Galactic field, but is consistent with the RM anomaly observed by
\citet{clegg}.  The overall size of this region is not well
constrained by the patchy point-source coverage, but it is at least as
big as CTB104A and may be as large as $\sim$ 8 degrees; such a large
scale feature must be taken into account when modeling the electron
distribution of the Galaxy.

We can make use of the observed variation of the RM to estimate the
electron density and the magnetic field strength in the shell of the
SNR.  Based on the flat spectrum, we can assume that the SNR has
already entered the radiative phase. Woltjer (1972) defined the
beginning of the radiative phase as the time where half of the
explosion energy is lost by radiation. The radius $R_{\rm rad}$ of the
SNR at this time depends on its explosion energy $E_{0}$ and the
ambient density $n_{0}$ before explosion:
\begin{equation} 
R_{\rm rad} = 23.1 ~E_{0}^{5/17} ~n_{0}^{-7/17}.
\end{equation} 
Assuming a typical explosion energy of 10$^{51}$ erg, we get a lower
limit for the ambient density $n_{0}\simeq2.0$.  The compression of
the material $\delta$ in the shell gives an electron spectrum, for
Fermi acceleration, with $\gamma$ defined by
\begin{equation}
\gamma = 2 - { {3 ~\delta}\over{\delta - 1}}.
\end{equation}
The spectral index $\alpha=-0.43$ found in Section~\ref{specan}, then
gives us a compression ratio of $\delta=4.8$, which is slightly higher
than the expected compression of 4.0 for an adiabatically expanding
SNR.  This results in a lower limit to the electron density of 
$n_{\rm e}= 9.6$ cm$^{-3}$, assuming the material in the shell is fully
ionized.
 
 The compression of $\delta=4.8$ and the linear radius of the SNR
$R=17.5$ pc give the width of the compression zone as 1.3 pc. This
corresponds to a shell thickness of only $\sim3\arcmin$, although the
total intensity map gives the impression of a thicker shell especially
towards the north. It is not surprising that the thin shell is not
noticeable in the radio map, because of projection effects. It should
be remembered that the remnant has not a perfect circular shape.  It
is in an advanced stage of evolution and the ambient medium structure
is quite inhomogeneous. Thus the expanding shell of the SNR should be
locally very thin, even though the radio image does not reveal it.

The rotation measure RM is given by:
\begin{equation}
RM = 0.81 ~n_{\rm e} ~B_{\|} ~L
\end{equation}
where $B_{\|}$ is the component of the magnetic field parallel to the
line of sight and $L$ is the path length through the compression zone
along the line of sight.  For a spherical geometry, $L$ can be
expressed in terms of the impact parameter $b$ as
\begin{equation}
L = \sqrt{R^2 - b^2} ~- ~\sqrt{r^2 - b^2}
\end{equation}
where $r$ is the inner radius of the compressed region.  The
tangential magnetic field configuration then leads to
\begin{equation}
B_{\|} = B_{0} ~{ {b}\over {(R+r)/2} },
\end{equation}
where $B_{0}$ is the constant magnetic field within the compression
zone. Inserting Eqns 4 and 5 into 3 gives
\begin{equation}
RM (b) = 1.62 ~n_{\rm e} ~B_{0} ~{ {b}\over {(R+r)} } 
       \left( ~\sqrt{R^2 - b^2} ~- ~\sqrt{r^2 - b^2} ~\right).
\end{equation} 
The rotation measure gradient at the center of the SNR with respect to
the impact parameter, $b$, would then be
\begin{equation}
 { {dRM(0)}\over{db}} = 1.62 ~n_{\rm e} ~B_{0} 
    ~{ {1-{r}/{R}}\over {1+{r}/{R}} }.
\label{fit}
\end{equation}
The observed RM distribution, as displayed in Figure~\ref{rmplot},
indicates that the gradient over the inner part of the remnant is
almost constant. We therefore can extract the gradient given in
Eqn~\ref{fit} by fitting a straight line to the RM distribution.  The
resultant fit gives a slope of $1.30 \pm 0.06$, also plotted in
Figure~\ref{rmplot}. Thus using the compression factor $\delta=4.8$,
corresponding to $r/R= 0.93$, and the lower limit for the electron
density $n_{\rm e}=9.6$ cm$^{-3}$ obtained above, we calculate the
constant magnetic field in the compressed region as $B_{0}\approx
2.3~\mu$G.  This would imply an ambient magnetic field strength of
about 0.5~$\mu$G.

\section{Summary}

We have presented new, high resolution \ion{H}{1} and continuum images
of CTB104A at 408 and 1420~MHz, augmented by CO spectral line maps and
HIRES IRAS data at 60$\mu$m.  Despite the superficial complexity of
CTB104A, a simple physical interpretation of the dominant processes at
work in the SNR emerges from the CGPS data.  We identify the
orientation of the local magnetic field, the physical origin of the
shell emission and the overall morphology of the surrounding neutral
shell that encloses the SNR.  The remnant as a whole is indeed
complex, however, as revealed by the signatures of localized
interaction between the remnant and the surrounding neutral medium.

Polarization and rotation measure structures are dominated by small
scale variations, induced by turbulence within the remnant.  However,
there is a distinct, ordered, rotation measure gradient from negative
values in the southeast to positive ones in the northwest.  The
direction of the rotation measure gradient defines the axis of the
magnetic field around the remnant, lying 30$\degr$ north of west. This
magnetic field is local to the remnant and generally counter to the
global magnetic field of the Galaxy in this area.  This magnetic field
structure lies inside a large RM anomaly inferred from rotation
measure data by \citet{clegg}.  The magnetic field strength within
CTB104A is $B_{0}\simeq2.3~\mu$G.

Non-thermal radio continuum emission arises from two shell edges at
the expected locations of compression for this magnetic field
configuration. A broken shell of neutral material surrounds the SNR.
One of the continuum shell edges, to the northeast, is broken in two.  At the
location of the break, the smaller of CTB104A's three prominences
extends from the shell.  We find no obvious disruption, or break, in
the surrounding neutral shell at the location of this smaller
prominence, probably due both to the smaller size of the prominence
and the overall complexity of the surrounding \ion{H}{1}
emission. Towards the two larger prominences, however, the neutral
shell is clearly broken, having been presumably less dense at these
points prior to disruption.  The morphology of the neutral medium, as
traced by \ion{H}{1}, is consistent with it being pushed by the
escaping ionized material in these regions.

In summary, the three prominences around the remnant reveal that
CTB104A is probably the best example of the leakage of blasts of
ionized material into a low density environment. 
One extension to the east is identified as a background \ion{H}{2} region
associated with a massive molecular shell.  CTB104A is a
relatively old supernova remnant, as deduced from its global spectral
index, expanding isothermally within a uniform magnetic field and
inside an associated \ion{H}{1} shell.  The distance 1.5 kpc inferred
from the radial velocity of the \ion{H}{1}, $v_{\rm LSR}\simeq-6$
km~s$^{-1}$, translates to a mean linear diameter of $\sim35$ pc.

\acknowledgments{
The Dominion Radio Astrophysical Observatory is a National Facility
operated by the National Research Council.  The Canadian Galactic
Plane Survey is a Canadian project with international partners, and is
supported by the Natural Sciences and Engineering Research Council
(NSERC). We wish to thank Dave Routledge for providing us with his CO
observations from 1994.  We thank Lloyd Higgs and Tom Landecker for
careful reading of the manuscript and discussions.}

\clearpage

\begin{figure}
\figcaption[f1.eps]{1420~MHz contour plot (top) and gray scale
image of  CTB104A with a resolution of
$65\arcsec \times 49\arcsec$ (EW $\times$ NS).  Contour
levels start from 4.8 K T$_{\rm B}$ and run in steps of 0.5 K T$_{\rm
B}$. The diffuse structure with a bright core in the lower left corner
is the \ion{H}{2} region S124. \label{i1420} }
\end{figure}

\begin{figure}
\figcaption[f2.eps]{408~MHz contour plot (top) and gray scale image
of  CTB104A with a resolution of $3\farcm7\times2\farcm8$
(EW $\times$ NS).  Contour levels start from 45 K T$_{\rm B}$ and run
in steps of 10 K T$_{\rm B}$.  The diffuse structure with a bright
core in the lower left corner is the \ion{H}{2} region S124
\label{i408}} 
\end{figure}

\begin{figure}
\figcaption[f3.eps]{Regions used for the TT-plot analysis are
displayed on a gray scale image of the region at 1420~MHz on a grid of
1 arc minute. The circles show the regions around the point sources
which were excluded.  The box with dotted lines shows the region used
for an overall spectral index calculation.  The dashed circle,centered
at $\ell=93\fdg7$ ad $b=-0\fdg3$, shows the approximate circular
part of the remnant and has angular dimensions of
$1\fdg2\times1\fdg2$.  The arrows indicate the positions of the
prominences extending away from the remnants boundary
\label{finder} }
\end{figure}

\begin{figure}
\figcaption[f4.eps]{Results of the TT-plot analysis for the
selected regions, numbered from 1 to 6. The spectrum relevant for
S~124 and 4C50.55 are given in the lower panel.  The lower-right panel
shows the spectra of the SNR obtained from the region plotted with
dotted lines in Figure ~\ref{finder}
\label{spec} }
\end{figure}

\begin{figure}
\figcaption[f5.eps]{Results of the TT-plot analysis towards the
four extensions of the SNR
\label{spec2} }
\end{figure}

\begin{figure}
\figcaption[f6.eps]{Polarized intensity image (top) of CTB104A at
1420~MHz with overlaid total intensity contours. Contours start at 7.0
K T$_{\rm B}$ and run in steps of 0.5 K T$_{\rm B}$.  Lower panel
shows the rotation measure image of the same region. White contour is
at $-$100 rad m$^{-2}$ and black contour is at 100 rad m$^{-2}$
\label{pirm} }
\end{figure}

\begin{figure}
\figcaption[f7.eps]{Rotation measure distribution as a
function of the angular distance from the center of CTB104A, ($\ell,
b$)= $93\fdg7, -0\fdg3$, projected on to the magnetic field axis,
which makes an angle of 30$\degr$ with respect to the Galactic plane.
The straight line is the fit to the RM gradient as explained in the
text
\label{rmplot} }
\end{figure}

\begin{figure}
\figcaption[f8.eps]{Neutral hydrogen data at 2$\arcmin$ resolution
integrated over 2 km~s$^{-1}$ intervals, between 10 and $-$6
km~s$^{-1}$. The overlaid contours are from the 1420~MHz continuum
image
\label{h1} }
\end{figure}

\begin{figure}
\figcaption[f9.eps]{Same as Figure~\ref{h1} but for
the velocity interval $-$8 to $-$24 km~s$^{-1}$
\label{h1b} }
\end{figure}

\begin{figure}
\figcaption[f10.eps]{
CO emission over the velocity interval between $-$55 and $-$35
km~s$^{-1}$ (left) and infrared 60 $\mu$m (right) images towards
CTB104A.  White contours show the total intensity emission at 1420~MHz
\label{hii} }
\end{figure}

\begin{figure}
\figcaption[f11.eps]{Direction of the magnetic field with respect
to the spiral arms in the Galaxy, obtained from the pulsar rotation
measure data, adopted from Han et al. (1999). The approximate position
of CTB104A is marked by a cross
\label{galrm} }
\end{figure}

\begin{figure}
\figcaption[f12.eps]{
Plot of the model  for an SNR expanding in a regular magnetic field.
The magnetic field direction and its parallel component given by the
pulsar rotation measure data are shown on the left. Also shown are the stretch
of the magnetic field lines due to the SNR expansion and the relative
RM strength and direction with respect to the given magnetic field
configuration.  The sketch on the right shows the same configuration
but the direction of the magnetic field is reversed 
\label{sketch}}
\end{figure}

\end{document}